\documentstyle[aps,epsf]{revtex}  

\input psfig
\tighten

\newcommand{\rt}{\rightarrow}
\newcommand{\ppj}{\psi^{'} \rightarrow
 \pi^+ \pi^- J/\psi}
\newcommand{\ppll}{\psi^{'} \rightarrow
 \pi^+ \pi^- J/\psi$, $J/\psi \rightarrow l^+ l^-}

\newcommand{\etal}{\it et al. \rm}

\newcommand{\bi}{\begin{itemize}}
\newcommand{\ei}{\end{itemize}}
\newcommand{\im}{\item}

\preprint{\vbox{\hbox{BIHEP-EP1-98-02\hfill}
                \hbox{UH511-904-98\hfill}
                \hbox{Submitted to ICHEP98}}}

\begin{document}        

\parindent=0.5 in
\baselineskip 14pt
\title{Recent Charmonium Results from  BES \cite{support}}
\author{Frederick A. Harris}
\address{University of Hawaii \\
Honolulu, HI 96822 }
%
\maketitle              
\vspace{-1.5in}
\begin{flushright}
UH511-927-99 \\
BIHEP-EP1-99-01 \\
hep-ex/9903036 \\
March 1999
\end{flushright}

\vspace{0.6in}
\begin{abstract}        
This paper summarizes recent results obtained from the BES $\psi(2S)$
data, which with $3.8 \times 10^6$ events, is the world's largest
data set.

\end{abstract}   	

\section{Introduction}               

The BES experiment runs at the Beijing Electron Positron Collider
(BEPC) which operates in the tau-charm energy range from 2 - 5 GeV.  
This paper is a review of recent results obtained from the $\psi(2S)$
data set, which is the world's largest sample.  Many details of this
work can be found in the references.

\section{The BES detector}

The Beijing Spectrometer, BES,
is a conventional cylindrical magnetic detector that is coaxial
with the BEPC colliding $e^+e^-$ beams.  It is 
described in detail in Ref.~\cite{bes}. A four-layer central drift
chamber (CDC) surrounding the beampipe provides trigger
information. Outside the CDC, the forty-layer main drift chamber (MDC)
provides tracking and energy-loss ($dE/dx$) information on
charged tracks over $85\%$ of the total solid angle.
The momentum resolution is $\sigma _p/p = 1.7 \% \sqrt{1+p^2}$ ($p$
in GeV/c), and the $dE/dx$ resolution for hadron tracks for this
data sample is $\sim 9\%$. 
An array of 48 scintillation counters surrounding the MDC provides 
measurements of the time-of-flight (TOF) of charged tracks with a resolution of
$\sim 450$ ps for hadrons. Outside the TOF system, a 12
radiation length lead-gas barrel shower counter (BSC),
operating in self-quenching streamer mode, measures the energies 
of electrons and photons over  80\% of the total solid
angle. The energy resolution is $\sigma_E/E= 22 \%/\sqrt{E}$ ($E$
in GeV).
Surrounding the BSC is a solenoidal magnet that
provides a 0.4 Tesla magnetic field in the central tracking
region of the detector. Three double layers of proportional chambers
instrument the magnet flux return (MUID) and are used to identify
muons of momentum greater than 0.5 GeV/c.

\section{\boldmath $J/\psi\rt\ell^+\ell^-$ 
branching fraction}

The branching fractions for the leptonic
decays $J/\psi\rightarrow e^+e^-$ ($B_e$) and $\mu^+\mu^-$
($B_\mu$) are basic parameters of the $J/\psi$ resonance.
In addition, these branching fractions are used
to determine the total number of $J/\psi$ events in a wide
variety of measurements that take advantage of the clean
experimental $J/\psi\rightarrow\ell^+\ell^-$ $(\ell=e$ or $\mu)$
signature. 

\begin{figure}[!htb]

\centerline{\epsfysize 2.5 truein
\epsfbox{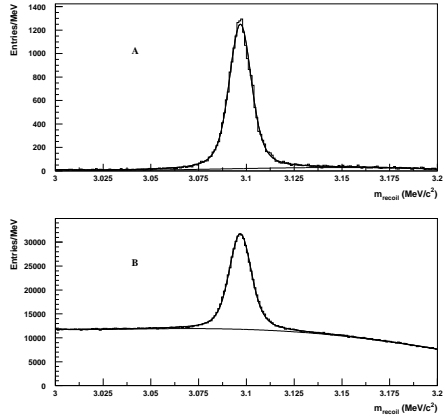}}
\caption{\label{fig:npsip} The $\pi^+\pi^-$ recoil
mass distributions for {\bf a)}
$\psi(2S) \rightarrow \pi^+ \pi^- J/\psi, J/\psi \rightarrow \ell^+\ell^-$
and {\bf b)} inclusive events.
}
\end{figure}

We determine the $J/\psi$ leptonic branching
fraction from a comparison of the exclusive and inclusive processes:
\begin{center}
\begin{tabbing} 
\hspace{50 mm}\= $\psi(2S) \rightarrow \pi^+\pi^-$ $J$\=$/\psi$           \\
              \>         \> $\hookrightarrow l^+l^- $\hspace{20 mm}\= $(I)$ \\
  and         \>         \> $\hookrightarrow$  anything \>$(II)$ 
\end{tabbing}
\end{center}
The $J/\psi$ leptonic branching fraction is determined from the relation
$$B(J/\psi\rightarrow l^{+}l^{-}) =
\varepsilon_{J/\psi} N^{obs}_{\ell}/\varepsilon_{\ell}
          N^{obs}_{J/\psi},$$
where $N^{obs}_{\ell}$ and $N^{obs}_{J/\psi}$ are observed numbers of events
for processes I and II (see Fig.~\ref{fig:npsip}), and
$\varepsilon_\ell$ and $\varepsilon_{J/\psi}$ are the respective acceptances.

The branching fractions are:
$B_e=(5.90 \pm 0.05 \pm 0.10)\%$ and
$B_{\mu}=(5.84 \pm 0.06 \pm 0.10)\%.$
The close equality of $B_{\mu}$ and $B_{e}$ is a verification
of $e$-$\mu$ universality:
$B_e/B_{\mu}=1.011 \pm 0.013 \pm 0.016. $
Assuming 
$B_{\mu}=B_{e}$, we find a combined leptonic branching fraction of
$B_l = B(J/\psi \rightarrow \ell^+\ell^-)=(5.87 \pm 0.04 \pm 0.09)\%$
and obtain a new world average
$B_l = (5.894 \pm 0.086)\%, $
which has an error about half that in the 1998 PDG\cite{pdg}.
This work  has been reported fully in Ref.~\cite{blldraft}.

\begin{figure}[!htb]
\centerline{\psfig{file=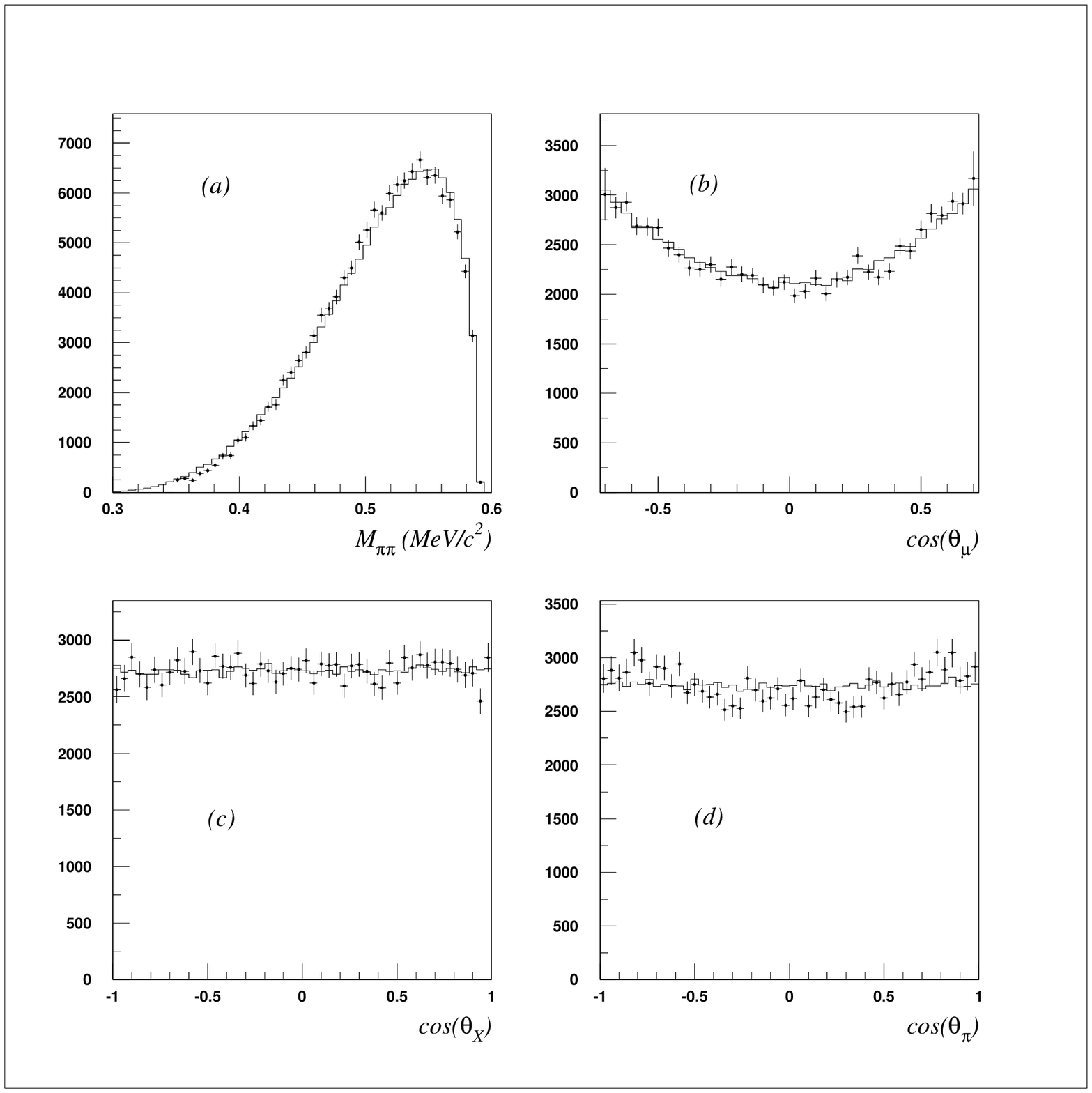,height=3.0in,bbllx=30
bp,bblly=160bp,bburx=532bp,bbury=656bp,rheight=3.0in,clip=}}
\caption{\label{fig:mccomp} 
Various distributions (corrected for detection efficiency)
for $\ppll$ decays. Dots with error bars are data; histogram is Monte Carlo
data.  {\bf a.)} $m_{\pi^+ \pi^-}$
distribution. 
{\bf b.)} $\cos \theta_l^*$ distribution. The
assumed distribution is a $1 + \cos^2 \theta_l^*$ distribution.
This angle is the angle between the beam direction and the $e^+$ in the
rest frame of the $J/\psi$.
{\bf c.)} $\cos
\theta_{X}$ distribution.  This is the cosine of the
angle of the $\pi \pi$ system with respect to the incoming $e^+ e^-$. 
{\bf d.)} $\cos
\theta_{\pi^+}^*$ distribution.  This is the cosine of the angle
of the $\pi^+$ with
respect to the $J/\psi$ direction in the $\pi \pi$ rest frame.
}
\end{figure}

\section{\boldmath $\psi(2S) \rt \pi^+ \pi^- J/\psi$}

The dynamics of the process $\ppj$, which is the largest decay mode of
the $\psi(2S)$, can be investigated using the very clean, high
statistics $\ppll$ events ($\sim$ 23 K).   This reaction may be pictured as
the radiation of two gluons by the quarkonium system as it transfers
to the lower energy level, followed by the hadronization of the gluons
into pions.
Early investigation of this
decay by Mark I \cite{abrams} found that the $\pi^+ \pi^-$ mass
distribution was strongly peaked towards higher mass values, in
contrast to what is expected from phase space.  Angular distributions
strongly favored S-wave production of $\psi \pi \pi$, as well as an
S-wave decay of the $\pi \pi$ system.

A comparison of our data and the Monte Carlo expectations based on the
results of Mark I is shown in Fig.~\ref{fig:mccomp}.  We find
reasonable  agreement except for the  $\cos \theta_{\pi^+}^*$
distribution, which is the cosine of the pion angle relative to
the $J/\psi$ direction in the $\pi \pi$ rest frame.  We find that
there is a D-wave
contribution in addition to the S-wave.

	One model that predicts a D-wave component is the
Novikov-Shifman model \cite{shifman}.  The pions in this process are
very low energy, so the process is a nonperturbative one.  This model uses
the scale anomaly and a
multipole expansion \cite{gottfried}-\cite{yan} to give an amplitude:

\begin{eqnarray}
A  \propto  \{q^2 - \kappa(\Delta M)^2(1 + \frac{2 m^2_{\pi}}{q^2}) 
+ \frac{3}{2} \kappa [(\Delta M)^2 - q^2](1 - \frac{4 m^2_{\pi}}{q^2})
(\cos^2 \theta_{\pi}^* - \frac{1}{3})\}, \nonumber
\end{eqnarray}
where $q^2$ is the four-momentum squared of the dipion system and
$\Delta M = M_{\psi(2S)} - M_{J/\psi}$.
The parameter 
$\kappa = (9/6 \pi) \alpha_s(\mu) \rho^G_{\mu}(\mu)$,
where $\rho^G$ is the gluon fraction of the $\pi$'s momentum, and
$\kappa$ is predicted to be $\approx 0.15$ to $0.2$.
The first terms in the amplitude are the S-wave part and the last is
the D-wave part.
Note that parity and charge conjugation invariance
require that the spin be even.
Fits using this amplitude are shown in Fig.~\ref{fig:jingyun}, and the
fit results are given in Table~\ref{table_kappa}.
Our result for $\kappa$ based on the $m_{\pi \pi}$
distribution is in good agreement
with that of ARGUS \cite{argus} using Mark I data for $\ppj$:  $\kappa =
0.194 \pm 0.010$ with a $\chi^2$/DOF of 38/24. Our results are preliminary.

\begin{minipage}[t]{3.2in}
\begin{figure}[!htb]
\centerline{\epsfysize 2.5 truein
\epsfbox{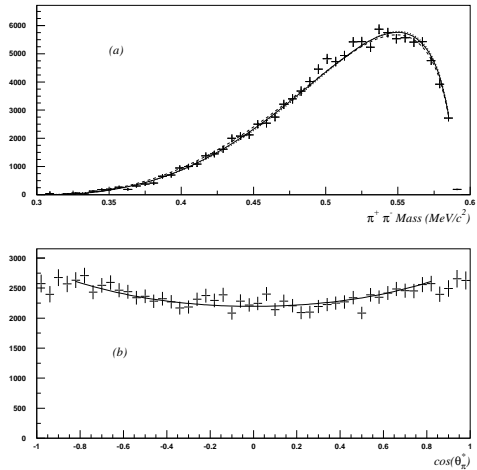}}
\caption{\label{fig:jingyun} 
Fits to 1D distributions.
{\bf (a)} $m_{\pi \pi}$
distribution.
{\bf (b)} $\cos
\theta_{\pi}^*$ distribution.
}
\end{figure}
\end{minipage} \ \
\begin{minipage}[t]{3.2in}
\begin{table}[!h]
\caption{Preliminary fit results for $\kappa$ for the Novikov-Shifman model.
}
\begin{center}
\begin {tabular}{lcc}
\label{table_kappa}
Distribution           &      $\kappa$     & $\chi^2/DOF$  \\ \hline
$m_{\pi \pi}$          & $0.186 \pm 0.003 \pm 0.010$ &  55/45     \\
$\cos \theta_{\pi}^*$    & $0.23^{+0.07}_{-0.04} \pm 0.11$ &  26/40     \\
$m_{\pi \pi}$ vs $\cos \theta_{\pi}^*$     & $0.183 \pm 0.003 \pm 0.005$ &  1618/1482     \\
\end{tabular}
\end{center}
\end{table}
\end{minipage}

\section{\boldmath Hadronic $\psi(2S)$ decays}
\noindent
Both $J/\psi$ and $\psi(2S)$ decays are expected to 
proceed via $\psi \rt ggg $, with widths that are 
proportional to the square of the $c \overline{c}$ 
wave function at the origin~\cite{appel}.  
This yields the expectation that
\begin{eqnarray}
\frac{B(\psi(2S) \rt X_h)}{B(J/\psi \rt X_h)} & \approx & \nonumber
\frac{B(\psi(2S) \rt e^+ e^-)}{B(J/\psi \rt e^+ e^-)} = (14.1 \pm 1.2) \%
\end{eqnarray}
It was first observed by MarkII\cite{mark2} that the vector-pseudoscalar (VP)
$\rho \pi$ and $K^*\overline{K}$ channels are suppressed 
with respect to the $ 14 \% $ expectation - the
``$\rho \pi$ puzzle''.
\newpage
\begin{table}[!h]
\caption{\label{hadronic} $\psi_{2S}$ Branching Ratios for Decays to Hadrons 
 (Preliminary).}
\begin{center}
\begin{tabular}{|c|c|c|c|}  
Channel  & $ {\cal B} (\psi' \rt X_h) (\times 10^{-4}) (PDG) $ & $ {\cal B} (\psi' \rt X_h) (\times 10^{-4})$ (BES) &  $ S = 0.14\frac{{\cal B}_{J/\psi}(PDG)}{{\cal B}_{\psi'} (BES)}$  \\\hline 
$\rho \pi$	& $ <0.83$  &	$ < 0.28$ & $> 64 $\\	
$K^+ \overline{K}^* (892)^- + c.c.$  
		& $<0.54$   &  $<0.30$ & $ > 23$ \\		
$K^0 \overline{K}^* (892)^0 + c.c.$  
		& -- &  $0.81 \pm 0.24 \pm 0.16$ & $ 7.3 \pm 2.7$ \\ 
$\omega \pi^0$	&   -- &  $0.38 \pm 0.17 \pm 0.11$ & $1.5 \pm 0.8$ \\		     
$\omega \eta$	&   -- &  $<0.33$ & $> 6.7 $\\		      
$\omega \eta^{'}(958)$ &   -- & $0.76 \pm 0.44 \pm 0.18$ & $0.3 \pm 0.2$  \\
$\gamma \eta         $  &   -- & $0.53\pm0.31\pm0.08$ & $2.3 \pm 1.4$      \\
$\gamma \eta^{'}(958)$  &   -- & 1.54 $\pm 0.31 \pm 0.20$ & $3.9 \pm 1.2$   \\\hline
$\omega f_2$	&   -- &  $<1.7$	& $> 3.5 $	    \\
$\rho a_2$	&   -- &  $<2.3$	& $ > 6.6$	     \\ 
$K^* (892)^0 \overline{K}^*_2 (1430)^0 + c.c.$ 
    &   -- &  $< 1.2$     			& $ > 7.8 $      \\ 
$\phi f^\prime_2 (1525)$ &   -- 		& $<0.45$	& $> 2.5$  \\

$\gamma f_2(1270)$    &  --  &  $3.01 \pm 1.12 \pm 1.07$ & $0.6 \pm 0.3$\\\hline
	
$K^* (892)^0 \overline{K}^* (892)^0$
	& -- &   $ 0.45 \pm 0.25 \pm 0.07$ & $< 1.6$\\ 
$\phi \phi$	& -- &  $ < 0.26$   &  \\ \hline
$b_1 \pi$   & -- & 5.3 $\pm$ 0.8 $\pm$ 0.8 &  $0.8 \pm 0.2$ \\ 
$K_1 (1270) \overline{K}$ 	& -- &   10.0 $\pm$ 1.8 $\pm$ 1.8 & $<0.41$  \\
$K_1 (1400) \overline{K}$ 	& -- &    $< 2.9$  &  $> 1.8$ \\ \hline
$\pi^+ \pi^- K^+ K^-$  & 16  $\pm$   4.0 & 6.9 $\pm$ 0.3 $\pm$ 1.2 & $1.5 \pm 0.5$ \\ 
     				
$K^+ K^- K^+ K^-$ 	&-- 	&   0.65  $\pm$ 0.10 $\pm$ 0.11 & $1.5 \pm 0.7$\\ \hline
$\pi^+ \pi^- \pi^0$	& 0.9  $\pm$ 0.5 & $ 1.06 \pm 0.11 \pm 0.16$ & $ 2.0 \pm 0.7 $\\  \hline

$K \overline{K} \pi$	& -- &   1.25 $\pm$ 0.18 $\pm$ 0.26 & $6.8 \pm 2.1$ \\  \hline

$K^* (892)^0 K^- \pi^+ \ c.c.$	& -- &  4.8 $\pm$ 0.5 $\pm$ 0.7 &  \\ 

$\phi K^+ K^-$		& -- & 0.51 $\pm$ 0.13 $\pm$ 0.09  & $4.0 \pm 1.4$ \\ 

$\phi \pi^+ \pi^-$      & -- & 1.3 $\pm$ 0.2  $\pm 0.2$ & $ 0.9 \pm 0.2$\\ 
$\omega \pi^+ \pi^-$    & -- & 4.7 $\pm 0.7 \pm 1.0$   & $2.1 \pm 0.5$ \\ 
$\rho^0 \pi^+ \pi^-$    &   4.2 $\pm$ 1.5 & 3.7 $\pm 0.6 \pm 0.9$ & \\\hline    
$\Lambda\overline{\Lambda}$ & $<4.0$  & 2.11 $\pm0.23\pm0.26$ & $0.9 \pm 0.2$\\ 
$\Sigma^0\overline{\Sigma^0}$ & -- & 0.94 $\pm0.30\pm0.38$ & $1.5 \pm 0.5$ \\
$\Xi\overline{\Xi}$ & $<2.0$ & 0.83 $\pm0.28\pm0.12$ & $1.5 \pm 0.6$ \\
$\Delta^{++}\overline{\Delta}^{--}$ & -- & 0.89 $\pm0.10\pm0.24$ & $1.7 \pm 0.5$   \\
\end{tabular}
\end{center}
\label{hbranching}
\end{table}
\normalsize
We have measured, as shown in Table~\ref{hadronic},
the $\psi(2S)$ branching fractions for
a large number of meson final states - many for the first time.
We confirm the $\rho \pi$ puzzle but with a bigger suppression factor and
note the following:
\begin{enumerate}
\im
The branching fraction for $\psi(2S)\rt \omega\pi^0$ 
is larger than the one for the isospin-conserving, $SU(3)$-allowed,
$\psi(2S)\rt\rho\pi$ decay.
\im 
Large isospin violations are seen between the 
branching fractions for charged and neutral 
$\psi(2S) \rt K^* \overline{K}$ decays, as shown in Fig.~\ref{fig:k*k}.
\im 
A clear signal is seen in $\psi(2S) \to \gamma\eta'$, 
but with a suppression factor of about four
with respect to the $14 \%$ expectation~\cite{etap}.
\im Decays to vector plus tensor final states, such as 
$\psi(2S)\rt \omega f_2, \rho a_2,
K^{*0} \overline{K_2^{*0}}$, and $ \phi f_2^{'} $ are also suppressed,
with suppression factors relative to the $J/\psi$ 
of at least about three or more \cite{vecten}. 
This is the first evidence for suppression in a channel other than VP.
\im
The $\psi(2S)\rt AP$ decays $b_1^+\pi^-$ and 
$K_1^+(1270)K^-$ are  
relatively strong decay channels of the $\psi(2S)$ \cite{prl}.
For $\psi(2S)\rt b_1^+\pi^-$, the result is higher than, but
consistent with what is expected from the 14 \% rule.
See Fig.~\ref{test}.
This rules out explanations for the $\rho \pi$ puzzle that suppress
all $\psi(2S)$ decays to lowest lying two-body meson final states.
\end{enumerate}
A summary of the models~\cite{blt} - 
\cite{chai},
attempting to solve 
the $\rho \pi$ puzzle, and related BES results, 
are reported in Table~\ref{models}.
\clearpage
\begin{minipage}[t]{3.2in}
\begin{figure}[!htb]
\centerline{\epsfysize 2.5 truein
\epsfbox{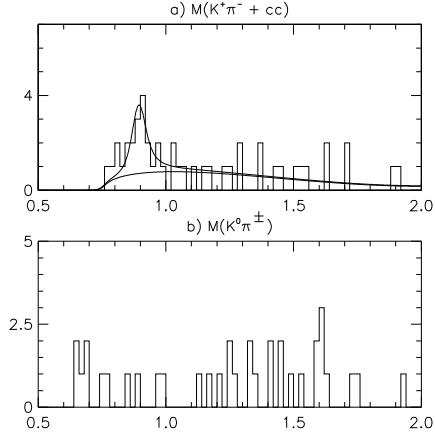}}
\caption{\label{fig:k*k} Invariant
{\bf a)} $K^{\pm} \pi^{\mp}$ and
{\bf b)} $K^o \pi^{\pm}$ mass distributions.
}
\end{figure}
\end{minipage} \ \
\begin{minipage}[t]{3.2in}
\begin{figure}[!htb]
\centerline{\epsfysize 2.3 truein
\epsfbox{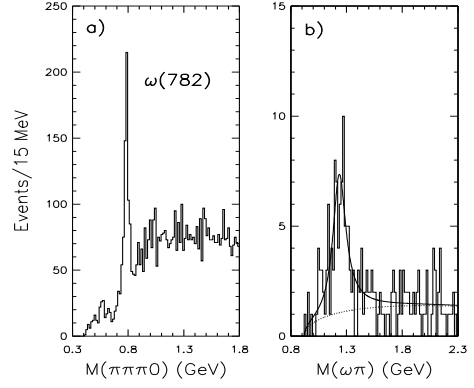}}
\caption{\label{test} 
{\bf a)} The $ \pi^+ \pi^- \pi^0 $ and
{\bf b)} the $\omega \pi^{\pm} $ mass distributions
from $\psi(2S) \rt \pi^+ \pi^- \pi^+ \pi^- \pi^0 $
events.  The $b_1$ is clearly visible in {\bf b)}.
}
\end{figure}
\end{minipage}

\nopagebreak

\begin{table}[!h]
\caption{Theoretical models related to the ``$\rho \pi$ puzzle'' and BES results.}
\begin{center}
\begin{tabular}{|c|c|c|}
Authors      & Model predictions &  BES results \\ \hline
Brodsky, Lepage, Tuan~\cite{blt}	& Hadron Helicity Conservation & $\psi(2S) \rt \omega \pi^0$ not suppressed. \\
 (1981, 1987) & suppresses $\psi(2S)$  & $\psi(2S) \rt VT$ are suppressed \\ 
  		& and $J/\psi \rt VP$ &  (these do not  violate H.H.C.) \\
                &                     &                       \\
Freund and Nambu~\cite{freund} (1975) & $J/\psi$-glueball ($ o $) mixing & $\mid m_o - m_{J/\psi}\mid < 80 MeV $ \\
Hou and Soni~\cite{housoni} (1983) &  explains $J/\psi \rt VP$ enhancement  & $ 4 MeV < \Gamma_o < 50 MeV $  \\\hline
Chen and Braaten~\cite{color}  & color-octet $ c {\overline c}$ production (requires H.H.C): 	&  \\ 
(1998)	&   $\psi(2S) \rt \omega \pi^0$ not suppressed	&  $\psi(2S) \rt \omega \pi^0$ not suppressed  \\\hline 
Brodsky and Karliner~\cite{bk}	& intrinsic charm  component in the light mesons   & \\
 (1997)   & (requires H.H.C.)  &   $\psi(2S) \rt \omega \pi^0$ not suppressed \\\hline
Pinsky~\cite{pinsky}  &  $\psi(2S) \rt VP$ are hindered M1 transitions : &  \\ 
  (1990)     &  $\psi(2S) \rt \omega f_2$ not suppressed  &  $\psi(2S) \rt \omega f_2$ is suppressed \\
 & $\psi(2S) \rt \gamma \eta' < 1 \times 10^{-5}$  &  $\psi(2S) \rt \gamma \eta' \sim 1.5 \times 10^{-4}$\\ \hline
Li, Bugg and Zou ~\cite{bugg} & Final State Interactions: &  \\
   (1997)  & may suppress $\psi' \rt VT$ & $\psi' \rt VT $ suppressed\\\hline
Chaichian and Tornquist~\cite{chai}    & invokes a form factor to suppress  & $J/\psi \rt \ K_1(1400){\overline K}$,\\
 (1989) & all 2-body meson modes& $\psi(2S) \rt b_1 \pi$, etc.. not suppressed \\
\end{tabular}
\end{center}
\label{models}
\end{table}

\section{\boldmath Studies of $\chi_{c_{0,1,2}}$ decays}
\noindent
The large sample of $\psi(2S)$ decays permits studies of
$\chi_{c0,1,2}$ decays with unprecedented precision ($ \sim 1 \times
10^6 \chi$'s). Some theoretical papers of interest are given in
Refs.~\cite{bolz} - \cite{duncan}, and a summary of branching fractions
is given in Table~\ref{chi_tbl}.  Many branching fractions, like
$B(\chi_{c0} \rt p \bar{p})$ \cite{gammachi}, are measured for the
first time.  Using $\chi_{c0} \rt \pi^+ \pi^-$, we find
$\Gamma_{\chi_0}=14.3\pm 3.6$~MeV \cite{gammachi}, which is an
improvement on the PDG value ($14 \pm 5$ MeV) \cite{pdg} based on
two discrepant measurements.

We have studied $\chi_c \rt 4 \gamma$'s and measured preliminary branching
ratios for $ Br(\chi_{c0}\rightarrow\pi^0\pi^0)$,
$ Br(\chi_{c2}\rightarrow\pi^0\pi^0)$, $
Br(\chi_{c0}\rightarrow\eta\eta)$,
and place an upper limit on $ Br(\chi_{c2}\rightarrow\eta\eta)$.
Fig.~\ref{fig:mchi2pipi} shows clearly the  $\chi_{c0}$ and
$\chi_{c2}$ in the $\pi^o \pi^o$ invariant mass distribution.
We find
$Br(\chi_{c0}\rightarrow\eta\eta)/
         Br(\chi_{c0}\rightarrow\pi^0\pi^0)=0.73\pm0.31\pm0.24$, where
0.95 would be expected assuming SU(3) flavor symmetry.
\clearpage
\begin{minipage}[t]{3.2in}
\begin{figure}[!htb]
\centerline{\epsfysize 2.3 truein
\epsfbox{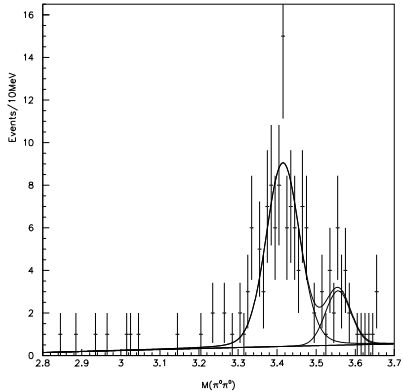}}
\caption{\label{fig:mchi2pipi} 
Invariant Mass of  $\pi^o \pi^o$. The $\chi_{c0}$
and $\chi_{c2}$ peaks are visible.}
\end{figure}
\end{minipage} \ \
\begin{minipage}[t]{3.2in}
\begin{figure}[!htb]
\centerline{\epsfysize 2.3 truein
\epsfbox{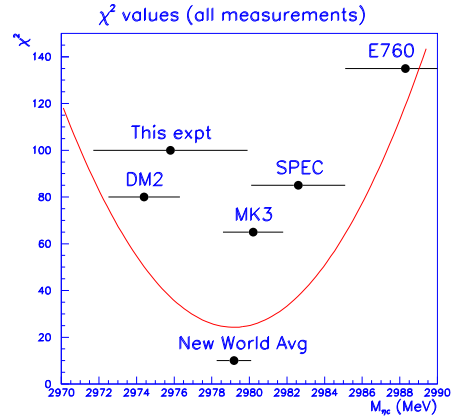}}
\caption{\label{fig:metac} 
$ M_{\eta_c}$ Results}
\end{figure}
\end{minipage}

\vspace{0.1in}
Using many decay modes of the $\chi_{c0}$, we have determined
$M_{\chi_{c0}} = 3414.1 \pm 0.6 \pm 0.8$ MeV.  This is a big
improvement over the PDG value which has an error of 2.8 MeV.
We have also measured the 
$\eta_c$ mass using $\eta_c \rt \pi^+ \pi^- \pi^+ \pi^-$, $\pi^+ \pi^-
K^+ K^-$, $K_s K^{\pm} \pi^{\mp}$, and $K^+ K^- K^+ K^-$,
and find
$M_{\eta_c} = 2975.8 \pm 3.9 \pm 1.2$ MeV.  Our result and the
previous ones are shown in Fig.~\ref{fig:metac}.  This work is
described fully in Ref.~\cite{metac}.  
Using $J/\psi$ data, BES finds a preliminary value of  $M_{\eta_c}$
that agrees well with the one obtained using the $\psi(2S)$ data.

\section{Summary}

Using the BES $\psi(2S)$ data set, which is the world's largest,
we have presented many results, including the
measurement $B(J/\psi \rightarrow \ell^+\ell^-) = 5.87 \pm 0.04 \pm 0.09$, 
many $\psi(2S)$ and $\chi_c$ branching ratios,
more information concerning the $\rho \pi$ puzzle, 
$M_{\chi_{c0}} = 3414.1 \pm 0.6 \pm 0.8~{\rm MeV}$, and
$M_{\eta_c} = 2975.8 \pm 3.9 \pm 1.2$ MeV.

I would like to acknowledge help from  Wei Yang (Colorado State
University) and Daniela Paluselli (University of Hawaii), who provided
some of the results presented in this paper.
I would also like to thank the staff of BEPC accelerator and the
IHEP Computing Center for their efforts.


\begin{table}[!p]
\caption{\label{chi_tbl}BES results on decay branching fractions of the
$\chi_{c0,1,2}$ charmonium states}

\begin{center}
\begin{tabular}{l|c|c}  
       &   BES  $(\times 10^{-3})$ &   PDG  $(\times 10^{-3})$ \\\hline 

$B(\chi_{c0} \to \pi^+ \pi^-)$
          &    4.68 $\pm$ 0.26 $\pm$ 0.65    &   7.5 $\pm$ 2.1 \\
$B(\chi_{c0} \to \pi^o \pi^o)$
          &    2.80 $\pm$ 0.32 $\pm$ 0.51    &  --  \\
$B(\chi_{c2} \to \pi^+ \pi^-)$
          &  1.49 $\pm$ 0.14 $\pm$ 0.22     &  1.9 $\pm$ 1.0   \\ 
$B(\chi_{c2} \to \pi^o \pi^o)$
          &  0.92 $\pm$ 0.27 $\pm$ 0.52     &  --   \\ \hline
$B(\chi_{c0} \to \eta \eta)$
          &  2.03 $\pm$ 0.84 $\pm$ 0.58     &  --   \\ 
$B(\chi_{c2} \to \eta \eta)$
          &   $<2.5$    &  --   \\ \hline
$B(\chi_{c0} \to K^+ K^-)$
          &    5.68 $\pm$ 0.35 $\pm$ 0.85   &   7.1 $\pm$ 2.4 \\
$B(\chi_{c2} \to K^+ K^-)$
          &  0.79 $\pm$ 0.14 $\pm$ 0.13    &  1.5 $\pm$ 1.1   \\ \hline
$B(\chi_{c0} \to p\overline{p})$
          &    0.159 $\pm$ 0.043 $\pm$ 0.053   &   $<0.9$ \\

$B(\chi_{c1} \to p\overline{p})$
          &  0.042 $\pm$ 0.022 $\pm$ 0.028     &  0.086 $\pm$ 0.012  \\

$B(\chi_{c2} \to p\overline{p})$
          &  0.058 $\pm$ 0.031 $\pm$ 0.032  &  0.10 $\pm$ 0.01   \\ \hline
$B(\chi_{c0} \to \pi^+ \pi^-\pi^+ \pi^-)$
          &    15.4 $\pm$ 0.5 $\pm$ 3.7  &   37 $\pm$ 7 \\

$B(\chi_{c1} \to \pi^+ \pi^-\pi^+ \pi^-)$
          &  4.9 $\pm$ 0.4 $\pm$ 1.2     &  16 $\pm$ 5  \\

$B(\chi_{c2} \to \pi^+ \pi^-\pi^+ \pi^-)$
          &  9.6 $\pm$ 0.5 $\pm$ 2.4    &  22 $\pm$ 5   \\ \hline
$B(\chi_{c0} \to K^0_s K^0_s)$
          &   1.96 $\pm$ 0.28 $\pm$ 0.52   &    -   \\
$B(\chi_{c2} \to K^0_s K^0_s)$
          &  0.61 $\pm$ 0.17 $\pm$ 0.16 &   - \\ \hline
$B(\chi_{c0} \to \pi^+ \pi^- K^+ K^-)$
          & 14.7 $\pm$ 0.7 $\pm$ 3.8  &   30 $\pm 7$ \\

$B(\chi_{c1} \to \pi^+ \pi^- K^+ K^-)$
          & 4.5 $\pm$ 0.4 $\pm$ 1.1 &   9 $\pm 4$ \\

$B(\chi_{c2} \to \pi^+ \pi^- K^+ K^-)$
          & 7.9 $\pm$ 0.6 $\pm$ 2.1  &   19 $\pm 5$ \\ \hline

$B(\chi_{c0} \to \pi^+ \pi^- p \bar{p})$
          & 1.57 $\pm$ 0.21 $\pm$ 0.54   &   5.0 $\pm 2.0$ \\
$B(\chi_{c1} \to \pi^+ \pi^- p \bar{p})$
          & 0.49 $\pm$ 0.13 $\pm$ 0.17  &   1.4 $\pm 0.9$ \\
$B(\chi_{c2} \to \pi^+ \pi^- p \bar{p})$
          & 1.23 $\pm$ 0.20 $\pm$ 0.35  &   -- \\  \hline

$B(\chi_{c0} \to K^+ K^- K^+ K^-)$
        &  2.14 $ \pm$ 0.26 $\pm$ 0.40  &    --  \\

$B(\chi_{c1} \to K^+ K^- K^+ K^-)$
        &  0.42 $ \pm$ 0.15 $\pm$ 0.12 &    --  \\

$B(\chi_{c2} \to K^+ K^- K^+ K^-)$
        &  1.48 $ \pm$ 0.26 $\pm$ 0.32 &    --  \\  \hline

$B(\chi_{c0} \to \phi  \phi)$
        &   0.92   $\pm$ 0.34 $\pm$ 0.38  &  --  \\

$B(\chi_{c2} \to \phi  \phi)$
        &   2.00   $\pm$ 0.55 $\pm$ 0.61  &  --  \\  \hline

$B(\chi_{c0} \to K^0_s K^+ \pi^- + c.c.)$
        &   $< 0.71$     &  --  \\

$B(\chi_{c1} \to K^0_s K^+ \pi^- + c.c.)$
        &   $2.46 \pm 0.44 \pm 0.65$    &  --  \\

$B(\chi_{c2} \to K^0_s K^+ \pi^- + c.c.)$
        &   $<1.06 $     &  --  \\ \hline
$B(\chi_{c0} \to 3(\pi^+ \pi^-))$
        &   $11.7 \pm 1.0 \pm 2.3$    &  $15 \pm 5$  \\
$B(\chi_{c1} \to 3(\pi^+ \pi^-))$
        &   $5.8 \pm 0.7 \pm 1.2$    &  $22 \pm 8$  \\
$B(\chi_{c2} \to 3(\pi^+ \pi^-))$
        &   $9.0 \pm 1.0  \pm 2.0$    &  $12 \pm 8$  \\ 
\end{tabular}
\end{center}
\end{table}

\end{document}